# Evolution of orbital angular momentum in a soft quasi-periodic structure with topological defects

Wang Zhang,[1] Jie Tang,[1] Peng Chen,[1] Guo-xin Cui,[1] Yang Ming,[2,*] Wei Hu,[1] and Yan-qing Lu[1,**]

[1]*National Laboratory of Solid State Microstructures, College of Engineering and Applied Sciences, Collaborative Innovation Center of Advanced Microstructures, Nanjing University, Nanjing 210093, China*
[2]*School of Physics and Electronic Engineering, Changshu Institute of Technology, Suzhou 215000, China.*

We propose a quasi-periodic structure (QPS) with topological defects. The analytical expression of the corresponding Fourier spectrum is derived, which reflects the combined effects of topological structure and quasi-translational symmetry. Light-matter interaction therein brings unusual diffraction characteristics with exotic evolution of orbital angular momentum (OAM). Long-range correlation of QPS resulted in multi-fractal and pairwise distribution of optical singularities. A general conversation law of OAM was revealed. A liquid crystal photopatterning QPS is fabricated to demonstrate the above characteristics. Dynamic reconfigurable manipulation of optical singularities was achieved. Our approach offers the opportunity to manipulate OAM with multiple degrees of freedom, which has promising applications in multi-channel quantum information processing and high-dimensional quantum state generation.

Topological features characterizing the global connection of a structure directly determine its physical properties. In past decades, many effects caused by variations in topological features, such as topological insulators [1,2] and the quantum Hall effect [3,4], have been observed. The topology of a structure can be changed by introducing a topological defect [5]. Such endowed global topology gives the structure the ability to reshape the wavefront of incident light to achieve orbital angular momentum (OAM) transition.

The light beam carrying quantized OAM is known as the optical singularity [6-8]. It has a singular topological phase distribution characterized by a phase winding factor $e^{il\phi}$. It has a donut-like intensity distribution that results from a phase singularity at the center of the optical axis. Owing to its unique topology, the optical singularity has quite important physical significance and practical applications, including micro-particle manipulation [9] and quantum information procession [10]. Many advances have been made in our understanding of the mechanisms of interaction for optical singularities with topologically modulated structures, both in linear [11-13] and nonlinear systems [14]. However, all of the studies to date have been limited to periodic structures. The evolution of OAM in quasi-periodic structures (QPS) with topological defects remain unknown.

Compared to ordinary periodic structures, QPS are deterministic disordered structures with long-range order [15-17]; they can be recognized as the intermediate state between periodic structures and completely disordered structures [18,19]. The transition from periodic structure to quasi-periodic structure undergoes the breaking of translational symmetry. This means that the conversation law of OAM must inevitably change during the transition according to Noether's theorem [20]; it is important to recognize and understand the effect of translational symmetry on diffraction characteristics. In addition, QPS have high-dimensional complexity, which may allow for OAM manipulation with multiple degrees of freedom. Therefore, there is a need to study the mechanism of the interaction between optical singularities and topologically modulated QPS, both in terms of physical meaning and practical application.

In this letter, we propose a one dimensional (1D) QPS with topological defects based on a soft reconfigurable liquid crystal (LC) consisting of alternative orthogonally planar aligned (PA) regions. Such structures were fabricated through photopatterning of a polarization-sensitive alignment agent with a dynamic microlithography system. We then observed the diffracted light field, which exhibited a localized optical singularity distribution with multifractal features, and formulated the conservation law of OAM. For comparison, we also studied the evolution of OAM in compound periodic structures (CPS) by turning structure parameters to reveal the effects of topological defects and translational symmetry on diffraction characteristics.

QPS can be obtained by the cut and project method [21,22]; that is, a restricted projection of a high-dimensional space (hyperspace) into a low-dimensional space. Therefore, the rank of the Fourier module of QPS equals the spatial dimension of the hyperspace. As a result, QPS have higher complexity than periodic systems. Accordingly, we constructed a 1D QPS as the 1D projection of a related two dimensional (2D) periodic lattice [18] and then modulated it with topological defects:

$$g(x) = sign\left\{\cos\left(\frac{2\pi}{\Lambda}x + l_1\phi\right) + \cos\left(\frac{2\pi\tau}{\Lambda}x + l_2\phi\right)\right\}, \quad (1)$$

where the slope of the projection axis $\tau$ is an irrational number, and $\Lambda$ is the pitch of the 2D periodic lattice. Moreover, such a structure is also considered as the incommensurately modulated structure derived from the irrational number. As a variable disordered parameter, a certain value of $\tau$ corresponds to a deterministic disordered QPS, which characterizes the translational symmetry of the system. In this work, the irrational golden ratio $(1+\sqrt{5})/2$ was chosen to construct the QPS without loss of generality. The $l_1$ and $l_2$ are topological charges that can have an integer or fractional value corresponding to different topological defects, and $\phi$ is the azimuthal angle. This topological modulation edits the global topology of the structure and endows it with quasi-angular-momentum at the same time; this has a unique effect on diffraction characteristics. Fig. 1(a) shows the structure of the QPS when the values of $l_1$ and $l_2$ are both taken as 1.

When light is diffracted by the 1D QPS with topological defects, its phase will be singularly modulated [23]. That means the incident light undergoes an OAM transition during the interaction. The distribution of the diffracted light field directly depends on the reciprocal lattice vector structure. Therefore, in order to analyze the mechanism of interaction, it is necessary to calculate the corresponding Fourier spectrum. This calculation process is cumbersome, while the result is given in the expression:

$$\begin{aligned}\mathcal{F}\{g(x)\} &= \sum_{m,n}\operatorname{sinc}\left(\frac{m+n}{2}\right)\operatorname{sinc}\left(\frac{n-m}{2}\right)e^{i(ml_1+nl_2)\phi}\delta\left(k - \frac{2\pi}{\Lambda}(m+n\tau)\right) \\ &= \sum_{m,n}f_{mn}e^{il_{mn}\phi}\delta(k-k_{mn})\end{aligned} \quad (2)$$

where $\text{sinc}(x) = \sin(\pi x)/(\pi x)$, $k_{mn} = \frac{2\pi}{\Lambda}(m+n\tau)$ is the reciprocal lattice vector corresponding to the diffraction order $(m,n)$, $l_{mn} = ml_1 + nl_2$ is the topological charge, and $f_{mn} = \text{sinc}\left(\frac{m+n}{2}\right)\text{sinc}\left(\frac{n-m}{2}\right)$ is the Fourier coefficient. Two indicators, $m$ and $n$, were needed to calibrate the diffracted light field for our 1D QPS, which is completely different from 1D periodic structures that have only one indicator. This allowed our QPS to achieve high-dimensional manipulation of optical singularities.

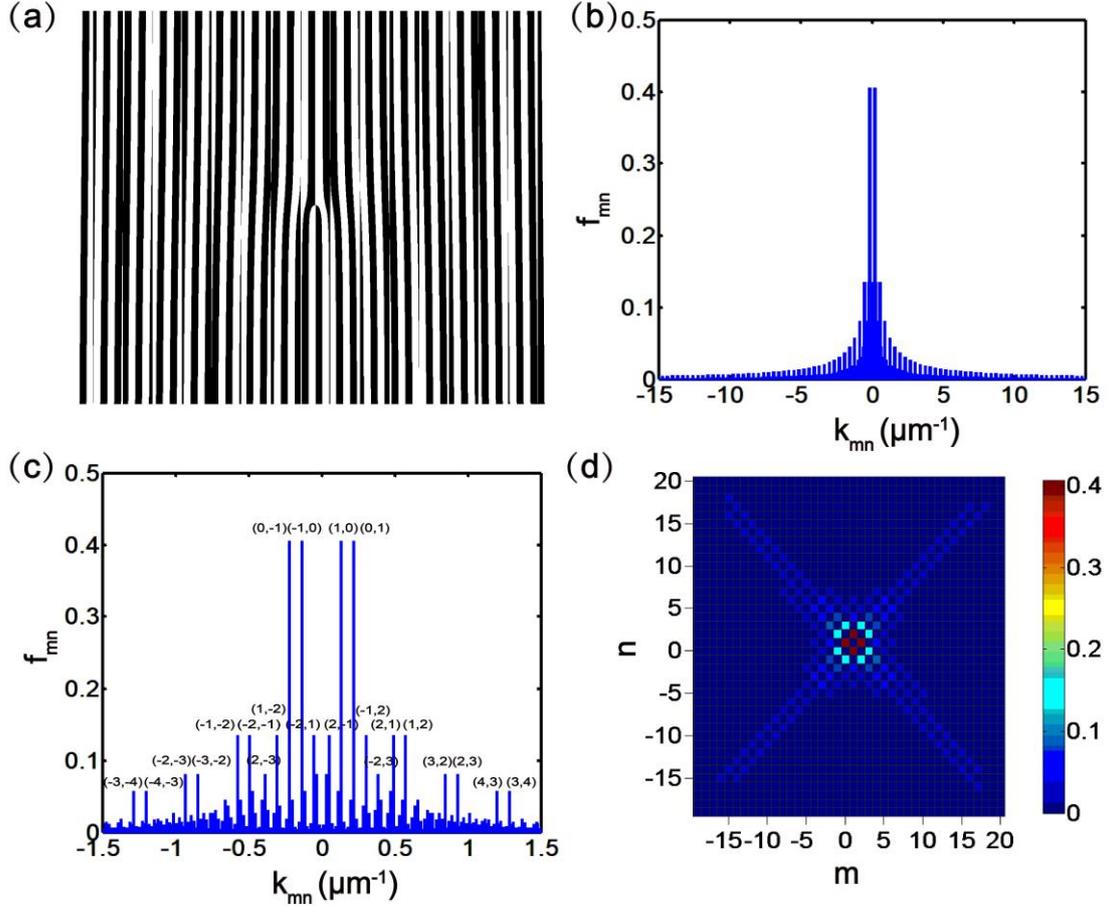

**FIG. 1.** Structure and Fourier spectra for QPS with $\tau = (1+\sqrt{5})/2$ and $l_1 = l_2 = 1$. (a) Structure of the QPS. (b) Self-similar Fourier spectrum of QPS for $\max(k_{mn}) = 15\,\mu\text{m}^{-1}$. (c) Fourier spectrum of the low-order diffraction part, with some diffraction orders labeled. (d) Two-dimensional distribution of the Fourier spectrum; the magnitude of the Fourier coefficient is represented by a color bar.

According to Fraunhofer diffraction theory, the intensity of far-field diffracted light is proportional to the square of the Fourier coefficient. Fig. 1(b) shows the Fourier spectrum when $k_{mn}$ is less than 15 μm$^{-1}$. Intuitively, the spectrum is self-similar, which is due to the long-range correlation of the structure [16,24]. This unique scaling feature can be analyzed by the so-called multifractal theory [25], which is usually used to characterize the statistical nature of a positive measure [26-28]. A positive measure is defined to describe a positive quantity that is distributed on the support of the measure. For our diffraction spectrum, the support was the reciprocal space and the positive quantity corresponded to Fourier intensity. To reveal the multifractal characteristics of the diffracted light field, the generalized dimension $D_q$ and associated singular spectrum $f(\alpha)$ were calculated according to the approximate scheme [27]. The value of $D_q$ corresponds to scaling exponents for $q^{th}$ moments of the measure; $f(\alpha)$ describes the relative strength of different dimensions $\alpha$. The results are shown in Fig. 2.

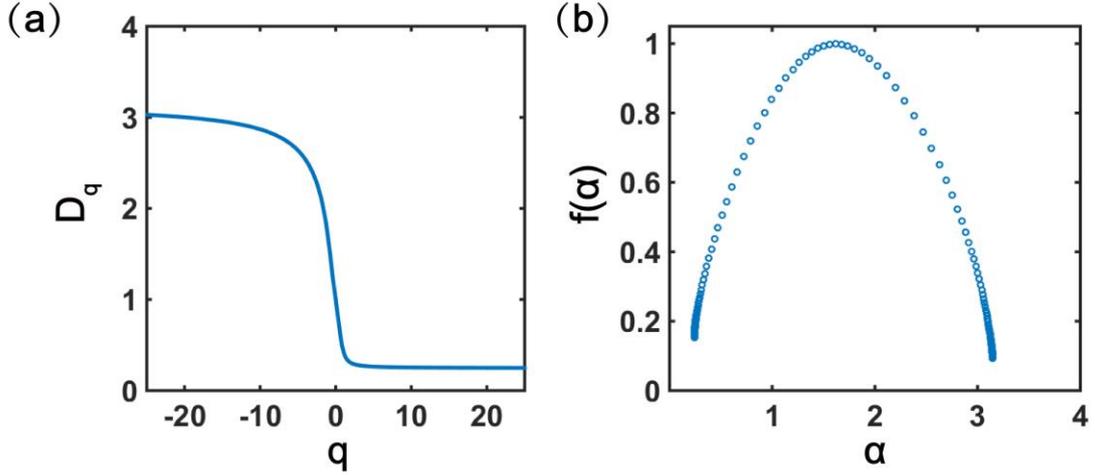

**FIG. 2.** Multifractal analysis of the Fourier spectrum for QPS. (a) Generalized dimension $D_q$. (b) Associated singular spectrum $f(\alpha)$. Maximum values of the Fourier coefficient subscript m (or n) and the order q for numerical calculation are 40 and 50, respectively.

The most rarefied and the most concentrated regions of the intensity measure were characterized by $\alpha_{max} = \lim_{q \to -\infty} D_q$ and $\alpha_{min} = \lim_{q \to +\infty} D_q$, respectively. The dimensions of $D_0 = f(\alpha_0) = 1$ arise from the whole real $k_{mn}$. The numerically calculated $f(\alpha)$ curve exhibited perfect smoothness, as shown in Fig. 2(b), which is the feature of an infinite structure. In Fig. 2(a), one can easily find the information dimension $D_1 < D_0$, which indicates that the intensity distribution of the diffracted optical singularities is a fractal measure.

It is noteworthy that optical singularities are characterized by the carried OAM [6]. Therefore the evolution of optical singularities directly corresponds to the numerical variation in OAM. In the interaction between the optical singularity and the proposed structure, the distribution of localized optical singularities with non-degenerate OAM was observed. Following the conservation of angular momentum, we identified a general conversation law of OAM in this system.

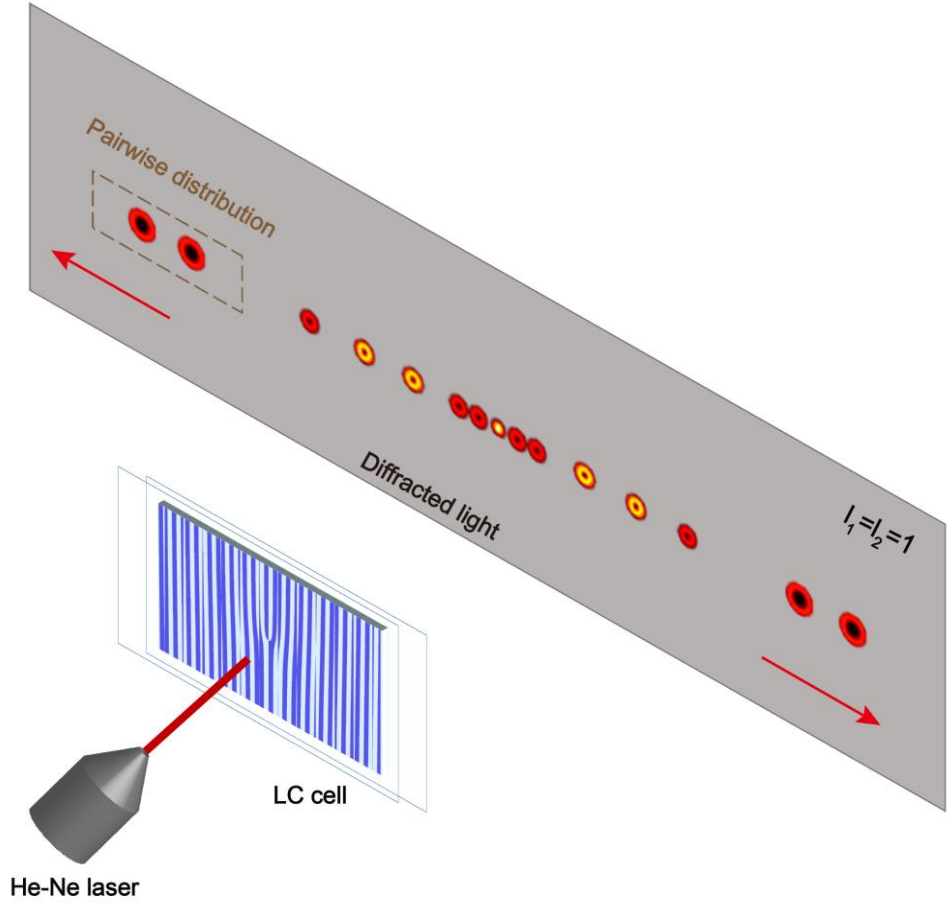

**FIG. 3.** Schematic illustration of a Gaussian beam emitted by a He-Ne laser incident on a reconfigurable photo-patterned liquid crystal (LC) cell. Optical singularities carrying OAM are generated and detected. Here we show the result of topological defects modulated with a QPS of $l_1=l_2=1$.

Firstly, we verified the non-degenerate nature of the reciprocal lattice vector structure of the QPS. Suppose that two diffraction orders, $(m,n)$ and $(m',n')$, are degenerate with each other, $k_{mn}$ must equal to $k_{m'n'}$. The slope of the projection axis can then be expressed by the form of a fractional number $\tau=(m'-m)/(n-n')$, which is contrary to the construction mechanism; therefore, for a QPS, it has a non-degenerate diffraction field distribution. That is, for any diffraction order, optical singularities carry only a single OAM value. That value is naturally derived from Eq. (2), expressed as：

$$l_{mn} = ml_1 + nl_2, \qquad (3)$$

Eq. (3) indicates the conversation law of OAM in a QPS with topological defects. The theoretical and experimental results of QPS with different topological defects are shown in Fig. 4. The diffraction orders were calibrated according to Fig. 1(c), and the corresponding OAM distribution was constrained to the conversation law. The optical singularities of any diffraction order carried a single OAM, as expected; these were regarded as localized states. The experimental results were in good agreement with theoretical simulations. Furthermore, it was noteworthy that the distribution of optical singularities was pairwise. Dotted rectangular frames in Fig. 4 mark two diffraction points' pairs at low order diffraction; the paired diffraction orders were:

$$(m, m-1) \text{ and } (m-1, m), \qquad (4)$$

Such an effect is directly derived from the characteristics of the pairwise distribution of the Fourier spectrum. The Fourier coefficient $f_{mn}$ is in the form of the product of two sinc functions, as expressed in Eq. (2). Thus, the non-zero component in the spectrum must satisfy $m+n = odd$ and $n-m = odd$ at the same time. Based on this, diffraction orders corresponding to the larger intensity should satisfy $n-m = \pm 1$, leading to the characteristics of a paired distribution. This feature can be understood more intuitively from the Fourier spectrum shown in Fig. 1 (d). For example, when m and n are both greater than 10, the pairwise distribution defined by the diffraction order above can be clearly seen.

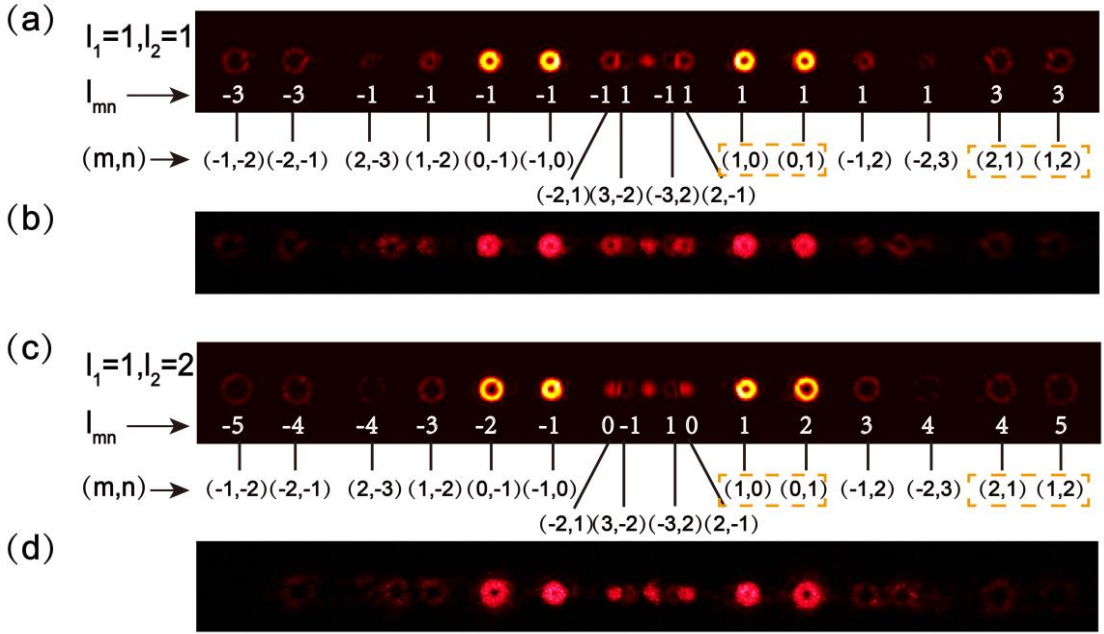

**FIG. 4.** Experimental and numerical distributions of diffracted optical singularities in QPS with various topological defects. The golden ratio $(1+\sqrt{5})/2$ is picked as the irrational number in all QPS. (a,b) Numerical and experimental results for $l_1 = l_2 = 1$. (c,d) Numerical and experimental results for $l_1 = 1$, $l_2 = 2$. Diffraction orders $(m,n)$ are labeled by brackets with two indices; the carried OAM of optical singularities are indicated correspondingly.

The 1D CPS was proposed to reveal the combined effects of topological structure and quasi-translational symmetry on diffraction characteristics and the evolution of OAM. In general, 1D CPS can be constructed in a way similar to the projection method; however, unlike the QPS, the slope of the projection axis here was a rational number instead. According to previous analysis, the reciprocal lattice vector structure of CPS must be degenerate. That means for any diffraction spot, it is actually composed of an infinite number of diffraction orders. If diffraction orders $(m,n)$ and $(m',n')$ are degenerate with each other, we can get the equation $k_{mn} = k_{m'n'}$ which can be easily expressed as $m + n\tau = m' + n'\tau$. The slope $\gamma$ corresponding to the degenerate axis on the Fourier spectrum was directly derived from that equation.

$$\gamma = \frac{n'-n}{m'-m} = -\frac{1}{\tau}, \qquad (5)$$

Fig. 5(a) and 5(b) show the Fourier coefficient spectra of the CPS with $\tau=2$ and $\tau=3/2$, respectively. The slopes of their degenerate axes were −1/2 and −2/3, as indicated by the green arrow, which is in full agreement with the theoretical analyses. One can imagine that topological charge corresponding to the degenerate reciprocal lattice vector structure should also be degenerate under ordinary circumstances. That means, a non-localized state with the hybrid optical singularity diffraction distribution shown in Fig. 5(c–f) will be observed under the transition of translational symmetry. However, when the structure parameters satisfied $l_2 = \tau l_1$, the transformation from the non-localized state into the localized state was observed as shown in Fig. 5(g–j). The reason for this special phenomenon is that optical singularities degenerated at the same diffracted spot carry the same value of OAM under such a condition. The locality transition of optical singularities demonstrates the combined effects of topological structure and translational symmetry.

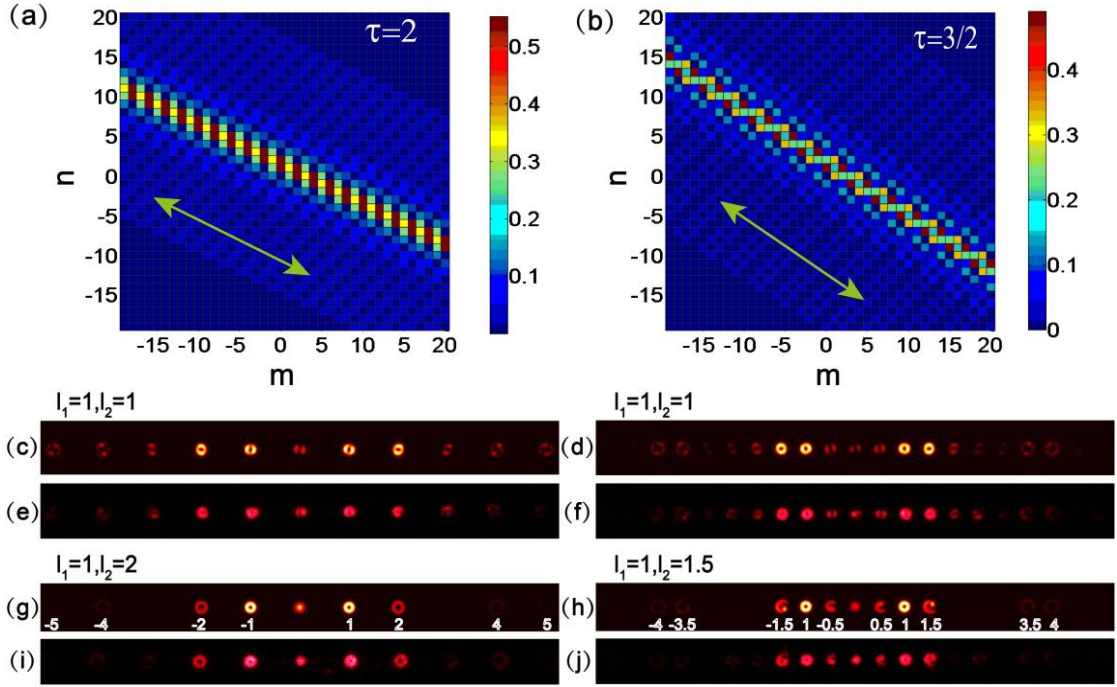

**FIG. 5.** Experimental and numerical distributions of diffracted optical singularities in CPS with various topological defects. (a,b) Fourier spectra for various CPS with $\tau = 2$ and $\tau = 3/2$, respectively; the degenerate axis is indicated by a green arrow. (c,e) Numerical and experimental results for an arbitrary topological defects doping case in CPS with $\tau = 2$. (g,i)

Numerical and experimental results under the specific condition $l_2 = \tau l_1$ in CPS with $\tau = 2$. (d–j) Analogy results of CPS with $\tau = 3/2$.

The photo alignment technique is a suitable way to achieve high resolution patterned domain LC alignments [29]. The designed LC orientations were performed through computer-controlled photopatterning via a dynamic microlithography system [12]. Here, we choose sulphonic azo-dye SD1 (DIC, Japan) as the alignment agent owing to its polarization-sensitive and rewritable advantages. LC directors are guided by the absorption oscillators of SD1 molecules, which are perpendicular to the incident UV light polarization [30]. Two pieces of glass substrates coated with SD1 were assembled to form a 6.0-μm thickness cell. Afterwards, we took a two-step photo exposure to achieve the adjacent orthogonal PA regions in order to avoid the influence of polarization sensitivity [11]. Finally, the designed structures were obtained by filling the cell with LC E7.

A linearly polarized 632.8-nm He-Ne laser illuminated the LC cell normally, as shown in Fig. 3. A CCD camera was used to capture the diffraction patterns. Fig. 4 shows the experimental results [Fig. 4(b,d)] and the compared numerical simulations [Fig. 4(a,c)] with different topological defects modulating QPS. The distribution of the diffracted optical singularities and the corresponding OAM are indicated clearly. In all cases, the obtained optical singularities were completely pure owing to the single carried OAM. This confirms the QPS can achieve high-dimensional localized modulation of optical singularities. In Fig. 5(a), Fourier spectra for CPS with $\tau=2$ are given. They have a degenerate axis with the slope $\gamma=-1/\tau$, as indicated by the green arrow. The corresponding experimental and theoretical results are shown in Fig. 5(e,i) and Fig. 5(c,g) respectively. The right column in Fig. 5 shows the juxtaposed results for CPS with $\tau=3/2$. Fig. 5(c–f) show that the diffracted optical singularities were all hybrid for an arbitrary modulated case, reflecting the transition of translational symmetry. However, when the structure parameters satisfied $l_2 = \tau l_1$, the non-localized optical singularities were transformed into localized ones, as shown in Fig. 5(g–j), reflecting the combined modulation of topological structure and translational symmetry.

We have mainly presented results for when the light beam carrying $l_0=0$ OAM was normally incident on the center of the structure. We also experimentally tested the case of $l_0 \neq 0$. The conversation law of OAM was then expressed as a generalized form: $l_{mn} = l_0 + ml_1 + nl_2$. When the incident position of the optical singularity did not coincide with the topological dislocation of the structure, coupling of spin and orbital angular momentum was taken into consideration. The evolution of OAM under such conditions was not intuitive and requires further investigation. Furthermore, other structured beams also deserve to be studied in this topological modulated QPS.

We choose LC as our experimental platform owing to its excellent electro-optical and soft reconfigurable properties. Based on photopatterning technology, LC offers the possibility to fabricate complex geometric phase structures. Yet, it is essentially a linear system. The interaction of an optical singularity with QPS or CPS in a nonlinear system such as a $LiNbO_3$ crystal can show more abundant physical features owing to the mix of OAM and frequency. For example, new quantum states [31] can be generated via the spontaneous parametric down conversion process.

Furthermore, a simple 1D QPS with 2D complexity is proposed here. Higher dimensional information can be compressed to 1D to construct a QPS according to our method. Moreover, the spatial dimension of the structure can also be extended to 2D and 3D. This may provide the possibility to achieve high dimensional and multi-degree-of-freedom regulation for OAM.

In conclusion, we proposed a 1D QPS with topological defects to reveal the effects of topological defects and translational symmetry on the evolution of OAM. The structural parameters were flexible owing to the soft reconfigurable LC crystal constructed using photopatterning technology. A general conversation law of OAM was revealed. We observed the distribution of diffracted optical singularities, which exhibited multifractal and pairwise features. We found that the transition of translational symmetry can lead to the non-localization of optical singularities. The transformation from non-localized state to localized state under $l_2 = \tau l_1$ was observed, which demonstrates the combined effects of topological structure and translational symmetry. Furthermore, these proposed structures with high-dimension complexity can manipulate OAM with multiple degrees of freedom, which is promising for multi-channel quantum information processing and high-dimensional quantum state generation.

This work was supported by the National Key R&D Program of China (2017YFA0303700), National Natural Science Foundation of China (Grant Nos. 61490714, 11604144, and 11704182), and Fundamental Research Funds for the Central Universities (14380001). We thank Prof. Chao Zhang for helpful discussions.

[*]Corresponding author.
*njumingyang@gmail.com*
[**]Corresponding author.
*yqlu@nju.edu.cn*